\documentclass[acmsmall,screen,review]{acmart}
\renewcommand\footnotetextcopyrightpermission[1]{} 
\AtBeginDocument{%
  }

\usepackage{footmisc}

\usepackage{amsmath}
\usepackage{algorithm}
\usepackage{algorithmic}

\usepackage{utfsym}

\usepackage{tabularx} 
\usepackage{array}

\usepackage{multirow} 
\usepackage{booktabs} 
\usepackage{makecell} 
\usepackage{array} 
\usepackage{multirow} 

\usepackage{ragged2e} 

\usepackage{caption}
\usepackage{array} 
\usepackage{cellspace} 
\usepackage{array} 
\usepackage{cellspace} 
\usepackage{colortbl} 
\definecolor{lightgray}{gray}{0.8} 

\usepackage[utf8]{inputenc}
\usepackage{listings}
\usepackage{xcolor}

\lstdefinelanguage{Solidity}{
    keywords={contract, function, public, payable, mapping, address, uint256, msg, sender}, 
    keywordstyle=\color{cyan!80!black}\bfseries, 
    morestring=[b]", 
    stringstyle=\color{orange!80!black}, 
    morecomment=[l]{//}, 
    morecomment=[s]{/*}{*/}, 
    commentstyle=\color{gray!60}\itshape, 
    sensitive=true 
}

\begin{document}
\title{From Data Behavior to Code Analysis: A Multimodal Study on Security and Privacy Challenges in Blockchain-Based DApp}



\author{Haoyang Sun}
\authornote{The two authors contributed equally to this research.}
\author{Yishun Wang}
\authornotemark[1]
\email{yishunwang@hainanu.edu.cn}
\affiliation{%
  \institution{Hainan University}
  \city{Haikou}
  \country{China}}

\author{Xiaoqi Li}
\affiliation{%
  \institution{Hainan University}
  \city{Haikou}
  \country{China}}
\email{csxqli@ieee.org}



\begin{abstract}
The recent proliferation of blockchain-based decentralized applications (DApp) has catalyzed transformative advancements in distributed systems, with extensive deployments observed across financial, entertainment, media, and cybersecurity domains. These trustless architectures, characterized by their decentralized nature and elimination of third-party intermediaries, have garnered substantial institutional attention. Consequently, the escalating security challenges confronting DApp demand rigorous scholarly investigation.
This study initiates with a systematic analysis of behavioral patterns derived from empirical DApp datasets, establishing foundational insights for subsequent methodological developments. The principal security vulnerabilities in Ethereum-based smart contracts developed via Solidity are then critically examined. Specifically, reentrancy vulnerability attacks are addressed by formally representing contract logic using highly expressive code fragments. This enables precise source code-level detection via bidirectional long short-term memory networks with attention mechanisms (BLSTM-ATT). Regarding privacy preservation challenges, contemporary solutions are evaluated through dual analytical lenses: identity privacy preservation and transaction anonymity enhancement, while proposing future research trajectories in cryptographic obfuscation techniques.
\end{abstract}



\maketitle


\fancyfoot{}
\pagestyle{plain} 

\section{Introduction}

The emergence of Bitcoin has catalyzed unprecedented advancements in blockchain technology, positioning it as a pivotal research frontier. This technological evolution has undergone three distinct evolutionary phases: 1) The Blockchain 1.0 era, epitomized by Bitcoin, established the cryptocurrency paradigm through decentralized monetary systems and payment mechanisms, albeit with limited industrial applications. 2) The Blockchain 2.0 phase, marked by Ethereum's introduction of smart contracts with Turing-complete scripting capabilities, witnessed expanded application scenarios across banking, insurance, securities, and trust sectors through sophisticated development ecosystems. 3) The current DApp epoch, where blockchain serves as foundational infrastructure enabling cross-industry credential authentication, while DApp function as value-transfer vectors through cross-industry value transfer protocols.

Blockchain technology, frequently hailed as the linchpin of the Fourth Industrial Revolution, derives its transformative potential primarily through DApp implementations – architecturally defined as distributed applications underpinned by blockchain consensus mechanisms and self-executing smart contracts. As the hallmark of Blockchain 3.0, DApp ecosystems critically influence the maturation trajectory of blockchain infrastructures. Crucially, DApp are projected to epitomize future blockchain-enabled socioeconomic frameworks, particularly through their capacity to mediate trustless transaction frameworks across decentralized autonomous organizations.

Although the concept of blockchain was proposed by Satoshi Nakamoto \cite{nakamoto2008bitcoin} (or a team) as early as 2008, the application of this technology in the real world has only been in use for a few years, and there are still a series of issues that cannot be ignored, such as privacy protection and controllable supervision, the inability to achieve decentralization, security, and scalability simultaneously, and the operational efficiency of blockchain itself. According to data from the blockchain security company PeckShield (Pai Dun), there were 177 blockchain security incidents in 2019, resulting in economic losses as high as 7.679 billion US dollars, an increase of about 60\% compared to 2018 \cite{radanliev2024rise}. According to incomplete statistics from the National Blockchain Vulnerability Database, the number of blockchain security incidents in 2020 reached 555, an increase of nearly 240\% compared to 2019, with economic losses amounting to 17.9 billion US dollars \cite{agarwal2024blockchain}. According to data released by the foreign research institution Chainalysis, the amount of cryptocurrency crime in 2021 reached 14 billion US dollars, an increase of 79\% year-on-year. The losses caused by cryptocurrency fraud cases reached 7.8 billion US dollars, an increase of 82\% year-on-year, and the losses from hacking theft cases were approximately 3.2 billion US dollars, an increase of 516\% year-on-year \cite{kovalchuk2024cryptocurrency}.

This paper will, based on existing research, collect data related to DApp, and combine the structural characteristics of blockchain to analyze the privacy protection and security of decentralized applications (DApp) on blockchain from two aspects: security and privacy protection.

The principal theoretical and methodological contributions of this paper are tripartite: 
\begin{itemize}
\item First, a multidimensional analytical framework is developed for deconstructing DApp' behavioral patterns through heterogeneous data fusion techniques.
\item Second, an innovative vulnerability detection paradigm is established by implementing Bidirectional Long Short-Term Memory networks with Attention mechanisms (BLSTM-ATT) to identify reentrancy vulnerabilities in Solidity-based smart contracts at source code granularity.
\item Third, a systematic theoretical framework for privacy preservation is formulated through dual-aspect analysis of identity anonymization protocols and transaction obfuscation mechanisms \cite{priya2024improved}, incorporating formal verification of zk-SNARKs implementations and quantitative assessment of differential privacy parameters.
\end{itemize}

\subsection{Related Work}
DApp, which are derived from underlying blockchain platforms, are decentralized applications running on top of smart contracts based on P2P peer-to-peer networks. The underlying blockchain technology provides them with trustworthy data recording. Through searching for relevant information on the network, it is found that research on DApp by related platforms mainly focuses on aspects such as application domains, application platforms, quality assessment, and DApp architectures \cite{heo2024blockchain,jimmy2024enhancing,ray2024blockchain,liu2024pricing,liu2024gastrace}, with corresponding data and behavior analyses. Research on the privacy protection and security of DApp is concentrated on the security and privacy protection of the underlying technology they employ, namely blockchain technology.

In recent years, research on the security and privacy protection of blockchain technology has been conducted. Bu et al. \cite{bu2025enhancing} investigated the security risks of blockchain systems, reviewed attack cases on blockchain systems, and analyzed the exploited vulnerabilities. Taylor P. J. et al. \cite{taylor2020systematic} conducted a systematic analysis of common blockchain security protocols. Singh S. et al.\cite{singh2021blockchain} provided a detailed analysis of potential blockchain security attacks and proposed existing solutions to these attacks. In addition, some researchers have offered insightful perspectives on vulnerability detection in smart contracts \cite{li2024cobra,bu2025smartbugbert,li2024stateguard,li2024defitail} .

Regarding privacy protection, Abdikhakimov and Islombek \cite{abdikhakimov2024interplay} introduced privacy protection mechanisms from three aspects: the blockchain network layer, the transaction layer, and the application layer. Zhang et al.\cite{zhang2024v2v} classified blockchain privacy protection technologies into address obfuscation, information hiding, and channel isolation and analyzed and compared the implementations of these three types of privacy protection technologies. Shen et al.\cite{shen2024privacy} categorized blockchain privacy into identity privacy and transaction privacy and analyzed the security issues associated with these two types of privacy.

In terms of smart contract security issues, Christof Ferreira Torres et al. \cite{torres2021confuzzius} proposed a hybrid simulator called CONFUZZIUS, which effectively identifies more bugs through constraint-solving and dynamic data analysis. Rao et al.\cite{rao2024scalability} proposed a transaction-based classification detection method for Ethereum smart contracts. Palina Tolmach et al.\cite{tolmach2021survey} proposed formal models and specifications for smart contracts, as well as methods for verifying such specifications.

\section{Background}

\subsection{Blockchain}
Blockchain is a decentralized distributed database composed of multiple interconnected blocks linked through cryptographic algorithms. Each block contains the hash value of the preceding block, transaction data, timestamps, and other relevant information \cite{mao2024scla}. By leveraging consensus algorithms, the blockchain network achieves a consensus mechanism, ensuring data synchronization and eliminating the possibility of data forgery by any single node, thereby enabling a trustless system. Through mutually agreed protocols and smart contracts, nodes interact and compete autonomously, ensuring the system operates independently without human intervention. Cryptographic algorithms enable any participant to query data records via public interfaces while preventing data modification or repudiation \cite{li2024guardians}. The chained structure facilitates efficient and rapid retrieval of transaction data, ensuring the traceability of data and transactions.
\subsection{DApp}
A DApp \cite{izaguirre2024decentralized} represents the integration of traditional applications (APPs) with blockchain technology, typically operating on a peer-to-peer (P2P) \cite{bukar2023peer,tariq2025peer} network. It serves as an enhancement and extension of conventional applications, with the key distinction being its decentralized nature. In DApp, participant information is either anonymous or protected, and operations are conducted through nodes on a peer-to-peer network. Smart contracts provide the foundational framework for decentralization, enabling trustless interactions between participants. From a practical perspective, a DApp can be succinctly conceptualized as a combination of smart contracts and traditional applications \cite{li2025scalm}, where smart contracts establish the prerequisites for decentralization. Structurally, DApp involves interactions between a front-end interface and users, as well as between smart contracts and the blockchain. This makes DApp publicly accessible programs that operate transparently on a network, leveraging blockchain's inherent properties of immutability and traceability.
\subsection{Smart Contract}
The concept of smart contracts \cite{wu2024comprehensive} was first introduced by Nick Szabo in 1994. He defined a smart contract as "a set of promises, specified in digital form, including the protocols within which the parties perform on these promises." Smart contracts operate on distributed ledgers and can execute, verify, and enforce complex behaviors of distributed nodes based on predefined rules, without the need for third parties, thereby achieving functions such as programming and information exchange.\par
Smart contracts are intelligent electronic contracts that transform contractual agreements into code, which is then deployed on a blockchain. Once deployed, the code is publicly accessible and immutable. When external conditions change, such as a breach or contract expiration, smart contracts automatically trigger the execution of predefined actions. This automation ensures transparency, efficiency, and trust in transactions, eliminating the need for intermediaries and reducing the potential for disputes.
\subsection{Ethereum}
Ethereum \cite{john2025economics} is a decentralized, open-source public blockchain platform with smart contract functionality. It features an integrated Turing-complete programming language, enabling users to develop decentralized applications (DApp) according to their specific requirements. As an application runtime platform, Ethereum ensures data transparency by making all data publicly accessible to nodes and immutable to third-party modifications. The Ethereum Virtual Machine (EVM) facilitates the execution and invocation of smart contracts. Additionally, Ethereum employs an account model, which reduces the cost of batch transaction processing, simplifies programming and development, and broadens the scope of application scenarios.
\subsection{Security Threats}
As an emerging application, DApp are increasingly recognized by enterprises and organizations for their underlying blockchain technology \cite{liu2025sok}, which features decentralization, tamper resistance, and traceability. However, DApp still face significant security threats. In this context, we analyze the security threats of DApp from two perspectives: smart contract security and privacy protection.
\subsubsection{Smart Contract Security}
\
\newline
Smart contract security is a critical component of DApp security. In 2016, vulnerabilities in the smart contracts of The DAO project led to the transfer of over 3.6 million Ether, resulting in losses exceeding USD 50 million \cite{dhillon2017blockchain}. This incident caused a temporary downturn in blockchain development.
\subsubsection{Privacy Protection}
\
\newline
In blockchain systems, privacy protection primarily focuses on identity and transaction information, which can be divided into identity privacy protection and transaction privacy protection. According to a report by The Record in April 2021, over 533 million Facebook users' personal information was leaked on a hacking forum \cite{popoola2024critical}. In June 2021, more than 700 million LinkedIn user data records were sold on a dark web platform. These incidents have drawn urgent attention to data security and privacy protection issues \cite{patel2024integration}.

\section{Methodology}
\subsection{Data Analytics for DApp}
The advantages of DApp stem from the underlying blockchain technology, which enables data ownership and value transfer. DApp facilitate inter-industry integration, ensure product controllability and traceability, reduce operational and development costs, enhance transaction security, and improve user experience. Due to their decentralized nature, DApp are increasingly valued and adopted by enterprises and organizations. However, security and privacy protection issues in blockchain technology and smart contracts may introduce new challenges to DApp security. We have collected the behavioral data of DApp and analyzed their distribution from the perspectives of platform, type, and smart contract. This analysis of DApp data behavior forms the basis for our subsequent analysis.
\subsubsection{Data Collection}
\
\newline
Data collection is essential for this study on DAPP security and privacy protection. This section introduces the dataset. We gathered DAPP-related data, including the number, types, platforms, and transactions, from four websites: State of the DApp, DAppReview, Top Blockchain DApp, and DApp.com, covering the period from April 2015 to February 2022. The data are presented in Table.\ref{table_1}, and we performed statistical analysis of DAPP data behavior accordingly.

\begin{table}[h]
\centering
\begin{tabular}{|l|c|c|}
\hline
\textbf{Platform} & \textbf{DApp Count} & \textbf{Smart Contract Count} \\ \hline
Ethereum          & 2935                & 4890                          \\ \hline
Klaytn            & 80                  & 316                           \\ \hline
EOS               & 331                 & 550                           \\ \hline
\textit{Steem}    & 79                  & 177                           \\ \hline
Hive              & 56                  & 105                           \\ \hline
POA               & 21                  & 51                            \\ \hline
\textit{xDai}    & 21                  & 58                            \\ \hline
Neo               & 24                  & 30                            \\ \hline
Obyte             & 17                  & 162                           \\ \hline
OST               & 2                   & 3                             \\ \hline
Loom              & 14                  & 33                            \\ \hline
GoChain           & 7                   & 17                            \\ \hline
\textit{Blockstack} & 24                & 0                             \\ \hline
TRON              & 88                  & 281                           \\ \hline
ICON              & 16                  & 36                            \\ \hline
NEAR              & 23                  & 21                            \\ \hline
BSC               & 189                 & 354                           \\ \hline
\textit{Moonriver} & 37                & 82                            \\ \hline

\end{tabular}
\vspace{8pt}
\caption{Platform DApp and Smart Contract Counts}
\label{table_1}
\end{table}
\vspace{-15pt}
We collected data on 3,964 DApp. Based on their application scenarios, we categorized these DApp into 21 classes, including gaming, gambling, and finance. The distribution of these categories is shown in Table.\ref{tab:dapp_distribution}.

\begin{table}[htbp]
\centering

\begin{minipage}{0.45\textwidth}
\centering
\begin{tabular}{|l|c|}
\hline
\textbf{Category} & \textbf{DAPP Count} \\ \hline
Games             & 680                 \\ \hline
Gambling          & 611                 \\ \hline
Social            & 415                 \\ \hline
Finance           & 386                 \\ \hline
Exchanges         & 264                 \\ \hline
Development       & 222                 \\ \hline
NFT               & 201                 \\ \hline
Def'i             & 192                 \\ \hline
Media             & 169                 \\ \hline
Wallet            & 138                 \\ \hline
Marketplaces      & 131                 \\ \hline
\end{tabular}
\end{minipage}
\hfill
\begin{minipage}{0.45\textwidth}
\centering
\begin{tabular}{|l|c|}
\hline
\textbf{Category} & \textbf{DAPP Count} \\ \hline
Governance        & 94                  \\ \hline
Security          & 87                  \\ \hline
Yield-farming     & 80                  \\ \hline
Property          & 81                  \\ \hline
Tools             & 61                  \\ \hline
Identity          & 48                  \\ \hline
Energy            & 34                  \\ \hline
Health            & 32                  \\ \hline
Insurance         & 20                  \\ \hline
Storage           & 18                  \\ \hline
\end{tabular}
\end{minipage}
\vspace{8pt}
\caption{DAPP Category Distribution}
\label{tab:dapp_distribution}
\end{table}
\vspace{-10pt}
In June 2018, EOS emerged as a new blockchain framework and gained attention for its scalability in decentralized applications. However, Ethereum's greater decentralization and longer application release history have led to the majority of DApp still being deployed on Ethereum. Therefore, we have collected and analyzed data on transactions, active users, and Ether from January 2018 to September 2020. Specific activity details are shown in Table.\ref{tab:ethereum_activity_data}.

\begin{table}[h]
\centering

\begin{tabular}{|l|c|c|c|}
\hline
 & \textbf{Transaction Volume} & \textbf{Active Users} & \textbf{ETH} \\ \hline
Games & 14734372 & 764798 & 300356 \\ \hline
Gambling & 9155186 & 466603 & 5037386 \\ \hline
Social & 667250 & 143384 & 6265 \\ \hline
Finance & 8415631 & 2091082 & 24311498 \\ \hline
Exchanges & 21377239 & 2153735 & 16068296 \\ \hline
Development & 1306938 & 311914 & 293079 \\ \hline
Media & 684475 & 231868 & 3300 \\ \hline
Wallet & 6990099 & 1395655 & 1004362 \\ \hline
Market & 2435308 & 190737 & 147025 \\ \hline
Governance & 330860 & 103686 & 1919 \\ \hline
Security & 2397611 & 773461 & 16899 \\ \hline
Property & 913607 & 106301 & 42736 \\ \hline
Identity & 353166 & 58768 & 4617 \\ \hline
Energy & 12870 & 7414 & 3237 \\ \hline
Health & 263 & 96 & 0 \\ \hline
Insurance & 5745 & 1969 & 0 \\ \hline
Storage & 1281920 & 574544 & 8 \\ \hline
High Risk & 4842315 & 1760754 & 8446005 \\ \hline
\end{tabular}
\vspace{8pt}
\caption{Ethereum Activity Data}
\label{tab:ethereum_activity_data}
\end{table}
\vspace{-18pt}
\subsection{Analysis Results}
Based on the data collected in Section 3.1, we performed preprocessing and conducted a multi-angle analysis of DApp data behavior. This analysis addresses key questions, such as the current state of DApp and which types are more investment-worthy. The findings enhance our understanding of blockchain and provide a data foundation for security analysis and privacy protection.
\subsubsection{Platform Distribution Data}
\
\newline
With the continuous development and improvement of blockchain technology, DApp have also been growing rapidly. The growing recognition of DApp's value has drawn increasing attention from organizations and enterprises. It should be noted that the quantity changes mentioned in the figure below refer to the current number of DApp, which is the difference between newly added and discontinued DApp.

\begin{figure*}[htbp]
    \centering
    \includegraphics[width=0.95\textwidth]{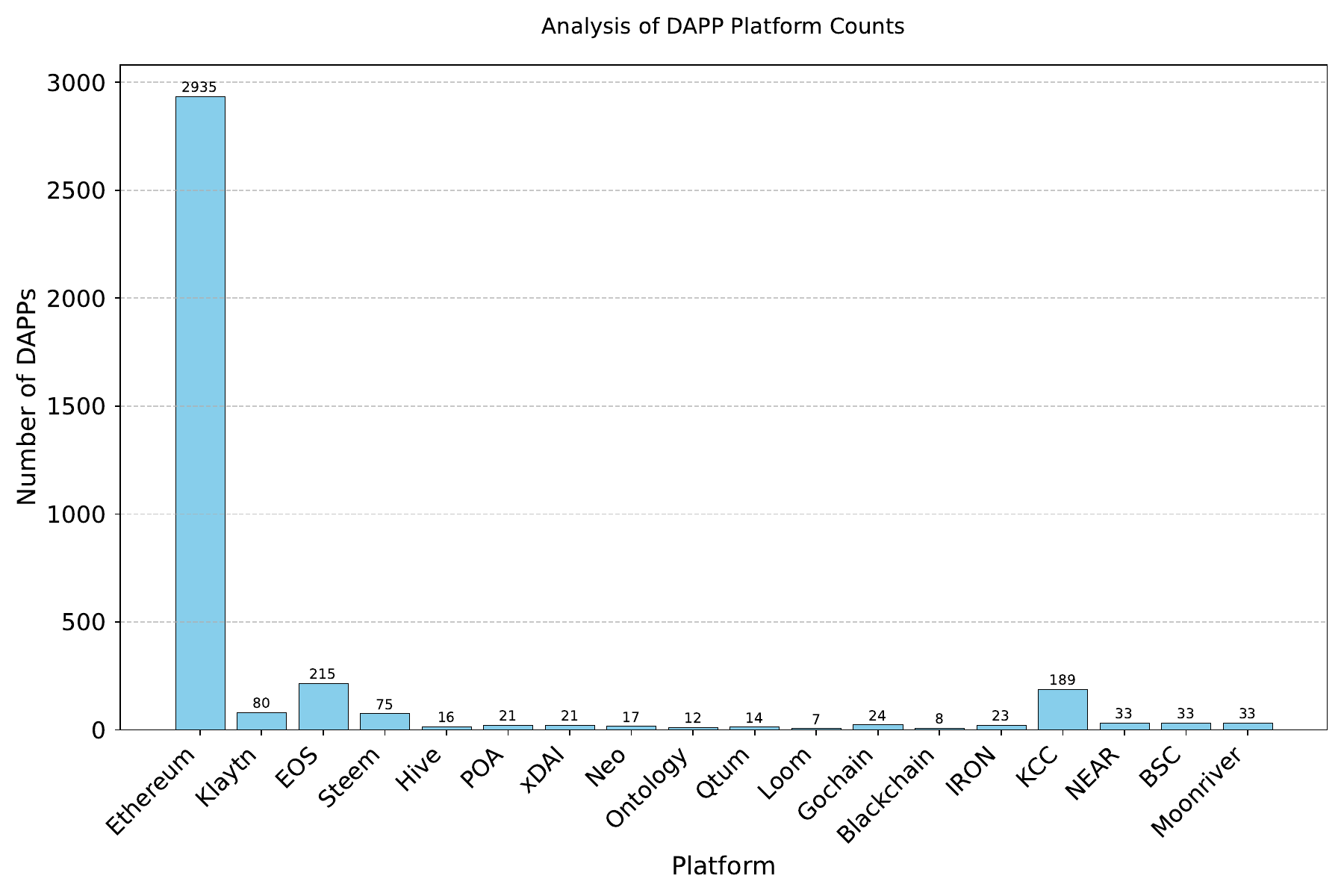}
    \caption{Analysis on the Quantity of DApp Platforms}
    \label{fig:1}
\end{figure*}

Currently, the leading blockchain development platforms for DApp are Ethereum, EOS, and TRON. As shown in Figure.\ref{fig:1}, Ethereum remains the dominant platform. While its longer existence contributes to this dominance, the data also reflect Ethereum's advantages in DApp development, with a remarkable share of 74.04\%.

Surveys indicate that the number of DApp deployed on Ethereum has consistently grown. Yet, the growth rate has shifted over time. From April 2015 to November 2018, the growth rate increased, but it has declined since November 2018. This change is mainly due to two factors: 1) Ethereum's security vulnerabilities, such as those in The DAO and Parity, which have raised risk concerns; and 2) the diversification of DApp development platforms, with the emergence of new platforms like EOS and TRON.

\subsection{Category Analysis}
\begin{figure*}[htbp]
    \centering
    \includegraphics[width=0.95\textwidth]{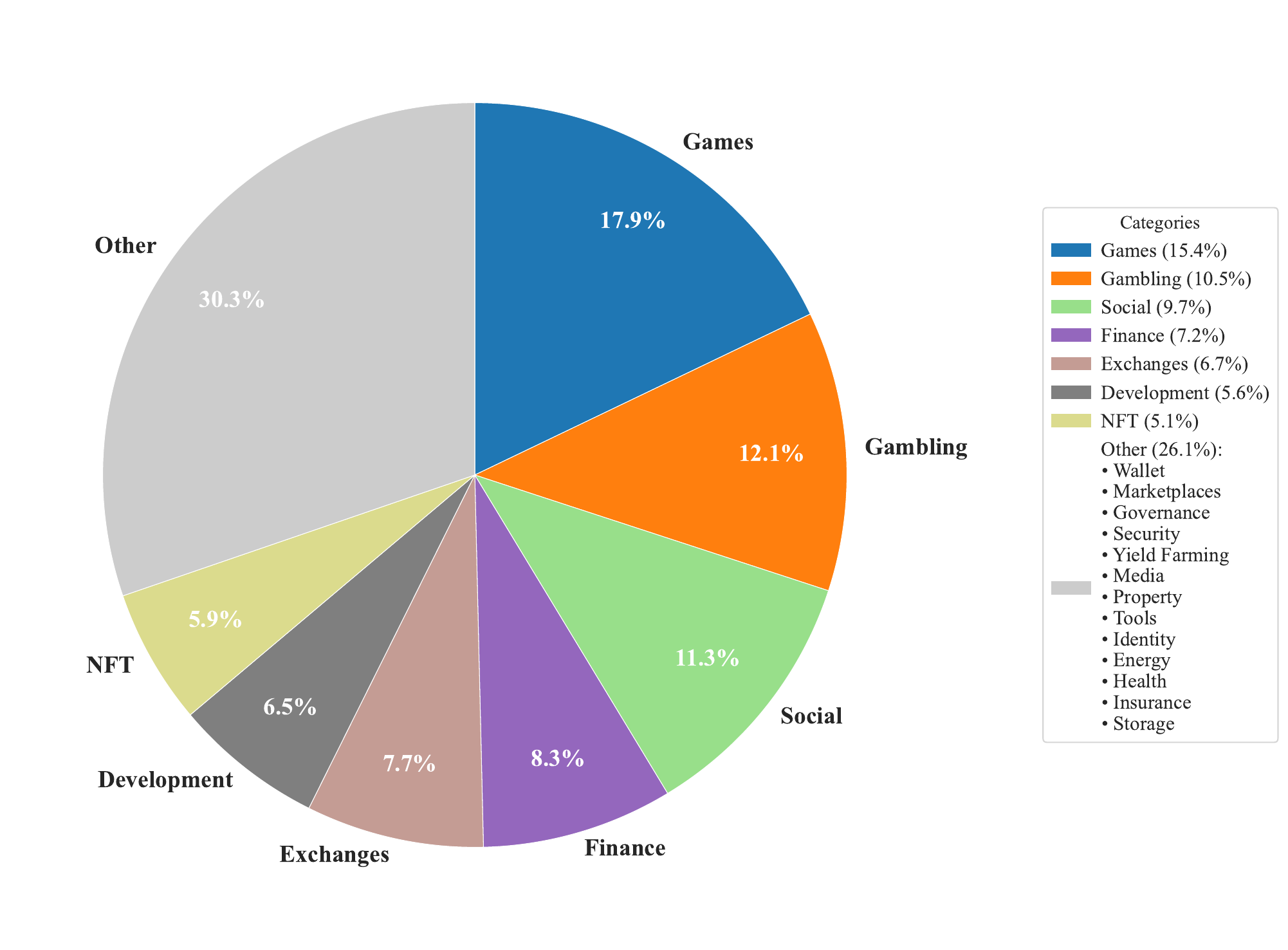}
    \caption{DApp Category Distribution}
    \label{fig:2}
\end{figure*}

The data in Figure.\ref{fig:2} show that five DApp categories—gaming, gambling, social, finance, and exchanges—account for 59.4\% of the total. Additionally, transaction volumes and active users, as indicated in Table 3.3.3 for Ethereum activity, are predominantly concentrated in these five categories.

From the above analysis, we can draw the following conclusions:
\begin{itemize}
    \item Finance-related DApp (finance and exchanges) are the most popular among users and have the highest number of active users, followed by entertainment-related DApp (gaming and gambling).
    \item In emerging fields such as health, insurance, and energy, the number of DApp, smart contracts, and active users is relatively limited.
\end{itemize}

\subsection{Reentrancy Vulnerability Detection for Smart Contracts}
In the DApp development and application environment, smart contracts, which handle various business logic, are receiving increasing attention for their security issues \cite{li2017discovering,wang2024smart,li2024detecting}. Notably, the attack on The DAO, which exploited a reentrancy vulnerability in a smart contract, caused a temporary downturn in blockchain applications \cite{niu2024unveiling}. This highlights reentrancy attacks as a significant security threat to smart contracts.\par
Smart contract security is crucial for ensuring the safety and privacy of decentralized applications based on blockchain technology. Building on our analysis of DApp data behavior in the previous chapter, this chapter delves into reentrancy vulnerability attacks in smart contracts. We conduct an in-depth study of their data behavior to achieve detection and identification of reentrancy attacks.\par

Interest in smart contract security is growing, with researchers working to identify vulnerabilities. Traditional analysis has relied on formal methods \cite{samreen2020reentrancy,xue2020cross}. Meanwhile, advances in deep learning have expanded the role of neural networks. The LSTM model, effective for sequence tasks like speech recognition \cite{kulkarni2021integration} and text prediction \cite{santhanam2020context}, is particularly noteworthy.\par
The method used in this chapter can secure DApp deployed on blockchain systems. The reasons are as follows:
\begin{itemize}
    \item Smart contracts are integral to blockchain and DApp, so researching their security is vital for understanding the security of blockchain and DApp.
    \item Reentrancy attacks on smart contracts can be detected and identified using deep learning methods. This allows for the detection of smart contracts with reentrancy vulnerabilities, thereby mitigating the security risks faced by DApp.
\end{itemize}

We focus on the most common and severe vulnerability in EVM-based smart contracts: reentrancy attacks. This vulnerability is exploited when a contract attempts to send Ether before updating its internal state. Specifically, reentrancy attacks can occur when a function creates an external call to an untrusted smart contract.\par
When an attacker transfers Ether to a smart contract address, it triggers the attack contract's fallback function \cite{kong2024characterizing}. Malicious code hidden in this function can activate reentrancy, causing repeated transfer operations.\par
In smart contracts, the fallback function is automatically triggered in two scenarios: 1) when a contract call is made but no matching function is found, it is called by default; 2) when the contract receives an Ether transfer, the fallback function can also be executed.\par
\begin{figure*}[htbp]
    \centering
    \includegraphics[width=0.88\textwidth]{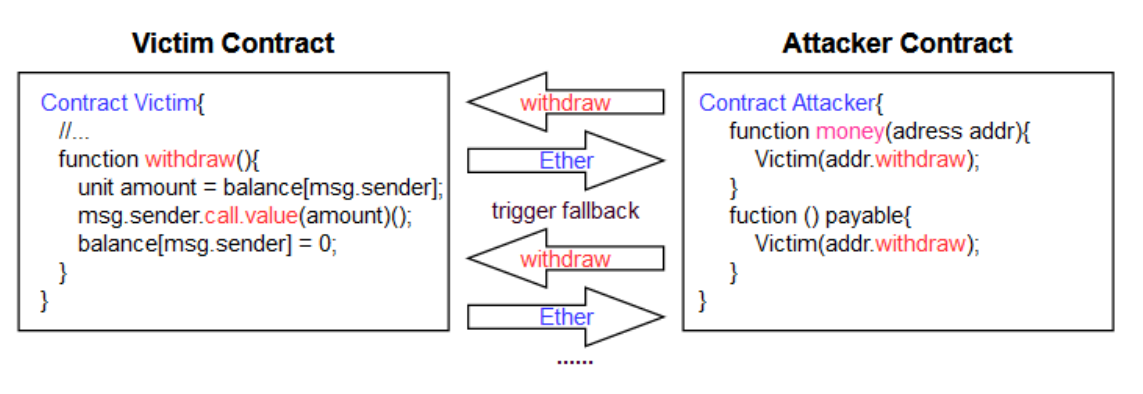}
    \caption{Reentrancy Attack Example}
    \label{fig:3}
\end{figure*}

In the example shown in Figure.\ref{fig:3}, the attack exploits the second trigger condition of the fallback function in a smart contract, as described earlier. The money function in the attack contract attempts to execute a withdrawal by calling the withdraw function of the victim contract. This action irreversibly activates the fallback function in the attack contract, which then repeatedly executes the withdrawal function until the Ether in the victim contract is depleted.

\subsubsection{Data Preprocessing}
\
\newline
Historically, there haven't been enough smart contracts to train neural networks. Today, with the increasing number of smart contracts across blockchain platforms, the time is right for neural network-based vulnerability detection. Deep learning methods are now employed to identify vulnerabilities in smart contracts.\par
\begin{figure*}[htbp]
    \centering
    \includegraphics[width=0.95\textwidth]{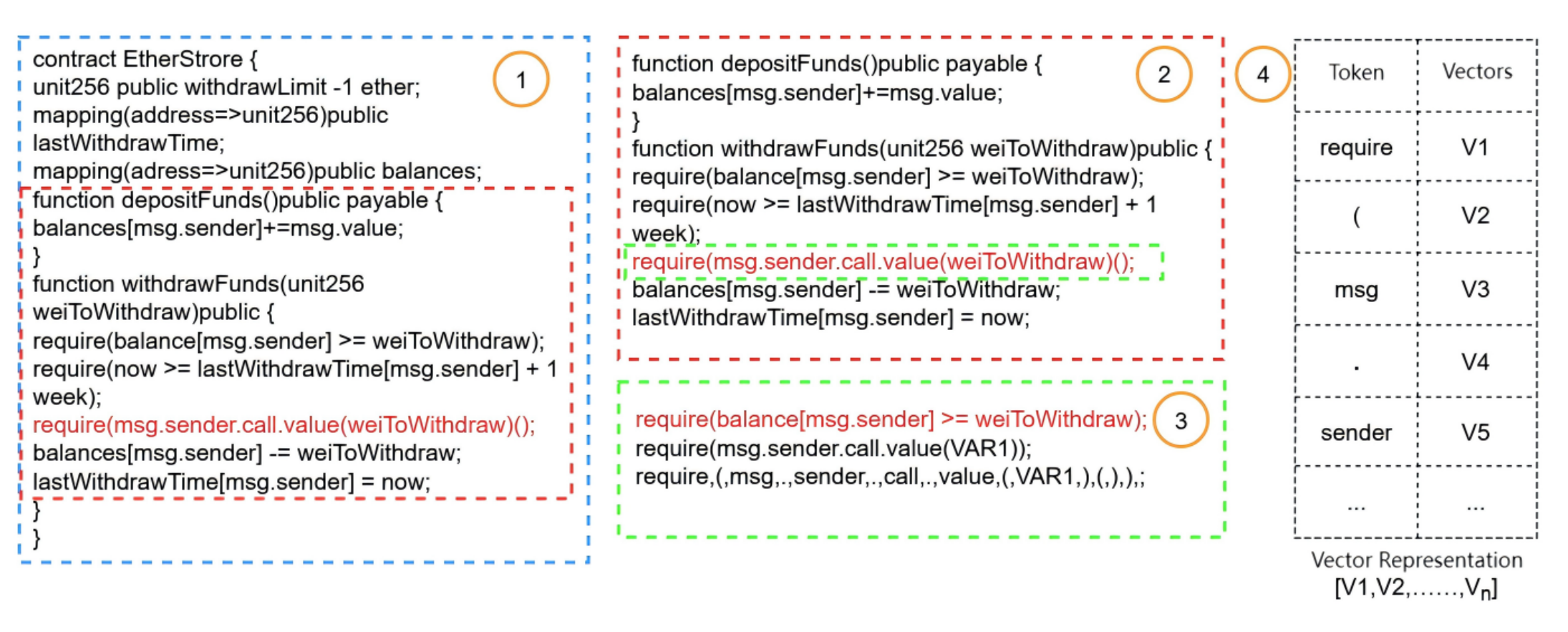}
    \caption{Data Processing Pipeline}
    \label{fig:4}
\end{figure*}

Our objective is to automatically determine if a given smart contract is reentrant using vulnerability detection methods. The automatic reentrancy vulnerability detection process involves several steps, as shown in Figure.\ref{fig:4}. First, data cleaning of the original smart contract is essential, such as removing blank lines, non-ASCII characters, and irrelevant comments. Then, the original smart contract is converted into contract snippets composed of key program statements. Next, each contract snippet is tokenized. Each snippet is then parsed into a series of code tokens, which are embedded into feature vectors for representation. Finally, during the experimental phase, these feature vectors are input into the adopted sequential model to train the detector, thereby achieving the detection of reentrancy vulnerability attacks.

Smart contracts on Ethereum are programs written in Solidity. They consist of multiple code lines, but some lines, like comments or unrelated functions, are irrelevant for reentrancy vulnerability analysis. To facilitate precise feature extraction, we condense smart contracts into expressive contract snippets.\par
Since deep neural networks typically use vectors as inputs, we need to represent smart contract snippets as vectors that are semantically meaningful for reentrancy detection. First, before generating vectors for each snippet, we obtain a symbolic representation through the following steps: 1) mapping user-defined variables to symbolic names (e.g., "VAR1," "VAR2"); 2) mapping user-defined functions to symbolic names (e.g., "FUN1," "FUN2"). After this, we perform a lexical analysis to split the symbolic representation of the contract snippet into a sequence of tokens.\par
Then, word2vec is used to convert these tokens into vectors. Word2vec maps tokens to integers and transforms them into fixed-dimension vectors. Since contract snippets may have varying numbers of tokens, the corresponding vectors can have different lengths. To ensure uniform vector length for input, vectors are padded with zeros at the end if shorter than the fixed dimension or truncated at the end if longer.\par

\subsubsection{Model}
\begin{figure*}[htbp]
    \centering
    \includegraphics[width=0.60\textwidth]{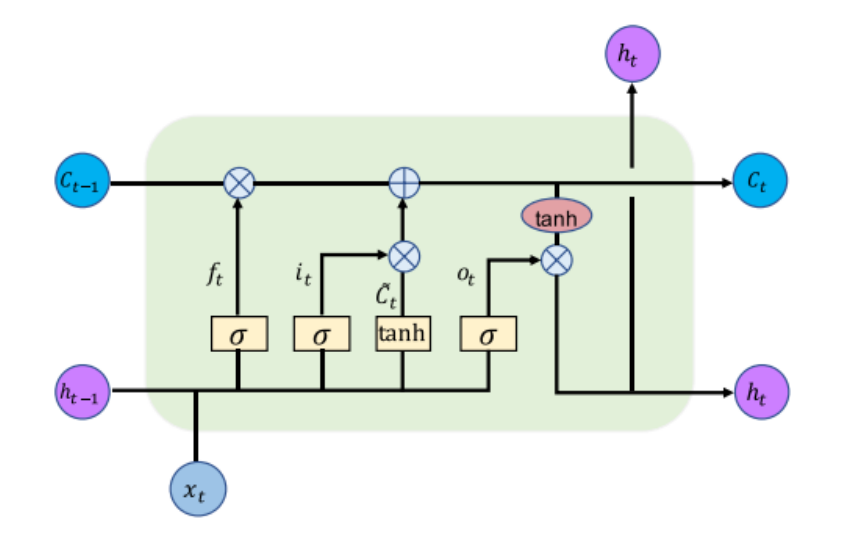}
    \caption{LSTM Unit}
    \label{fig:5}
\end{figure*}

\
\newline
An LSTM \cite{hosseinzadeh2025comprehensive,alizadegan2025comparative} unit consists of an input gate $i_t$, an output gate $o_t$, a forget gate $f_t$, and a cell state $C_t$, allowing the unit to remember values at any time and control information flow. As shown in Figure.\ref{fig:5}.

 The hidden state $h_t$ of an LSTM unit can be computed as follows:
\begin{equation}
f_t = \sigma(W_f \cdot[h_{t-1},x_t] + b_f)
\nonumber
\end{equation}
\begin{equation}
   i_t = \sigma(W_i \cdot[h_{t-1},x_t] + b_i) 
\end{equation}
\begin{equation}
   o_t = \sigma(W_o \cdot[h_{t-1},x_t] + b_o) 
\end{equation}
\begin{equation}
    \dot C_t = \tanh(W_c \cdot[h_{t_1},x_t] + b_C)
\end{equation}
\begin{equation}
    C_t = f_t \odot C_{t-1} + i_t \odot C_t
\end{equation}
\begin{equation}
    h_t = o_t \odot \tanh(C_t)
\end{equation}

In this context,$C_t$ represents a new candidate vector. The $Sigmoid$ function $\sigma$ and the hyperbolic tangent function $tanh$ are activation functions used within the LSTM unit. The symbol $\odot$ denotes matrix multiplication and element-wise multiplication. These functions share similar equations but differ in their parameter matrices $W$. Since standard LSTM cannot capture future information in a sequence, a bidirectional LSTM layer is added to address this limitation. \par
To highlight the importance of certain output results for vulnerability detection, we introduced an Attention Mechanism, resulting in the BLSTM-ATT \cite{zhifeng2022comparison,tereshchenko2023vulnerability} sequential model. For instance, for important words in lines of code (e.g., call.value), we use the Attention Mechanism to assign weights, which can be formalized as:
\begin{equation}
    \mu_t = \tanh(Wh_t + b)
\end{equation}
\begin{equation}
    \alpha_t = \frac{\exp(\mu^t\mu)}{\Sigma(\exp(\mu^T_t\mu))}
\end{equation}

\begin{figure*}[htbp]
    \centering
    \includegraphics[width=0.68\textwidth]{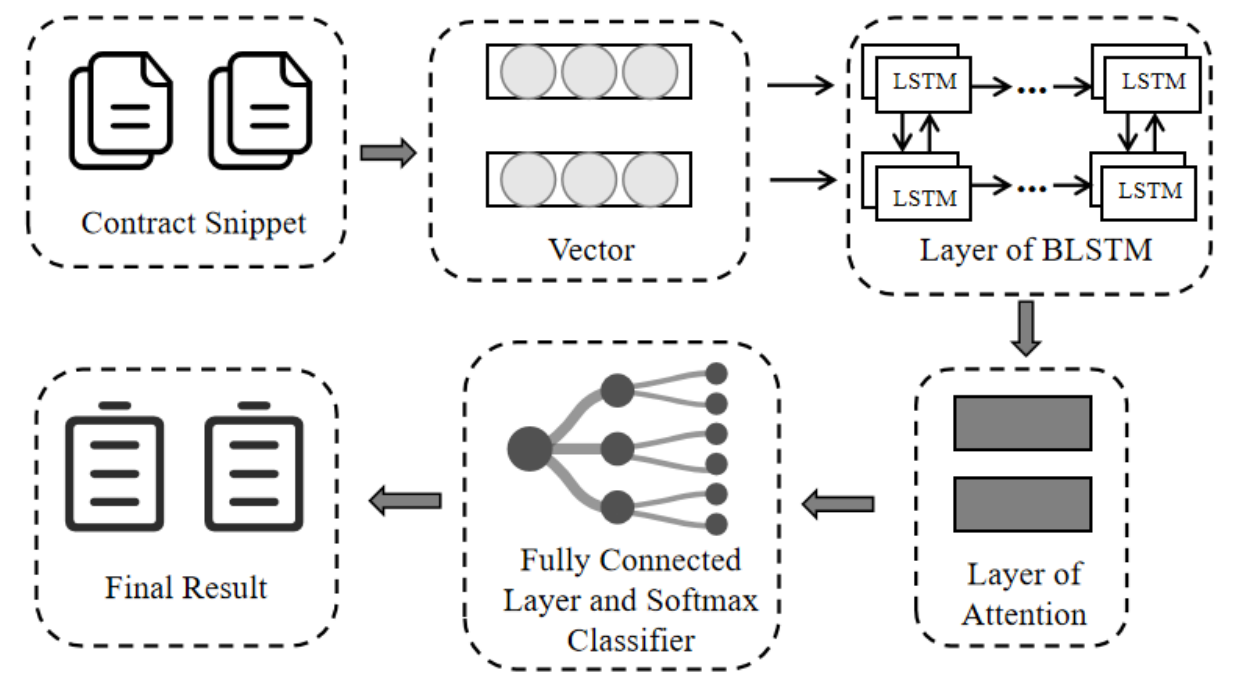}
    \caption{Model Framework}
    \label{fig:6}
\end{figure*}

$\alpha$ represents a normalized weight obtained through the Attention Mechanism. The specific model architecture is shown in Figure.\ref{fig:6}.

We input word2vec-generated feature vectors \cite{abubakar2022sentiment} into the BLSTM-ATT sequential model to learn model parameters. This involves calculating gradients and updating parameters during backpropagation. Once training is complete, the trained model is used for reentrancy detection.\par

Given one or more smart contract snippets from the test set, we convert them into vector representations and input these vectors into the sequential model. The model outputs results for each target smart contract, indicating whether it has reentrancy with "1" or "0".\par

Formally, we use a $softmax$ classifier to predict the label $y^*$ of the contract snippet S. The detector takes the hidden state $h_*$ as input:
\begin{equation}
    p(y/S) = solfmax(Wh^* + b)
\end{equation}
\begin{equation}
    y^* = argmax \cdot p(y/S)
\end{equation}

\subsubsection{Model Training Details}
In parameter settings, we use 10-fold cross-validation\cite{fushiki2011estimation} to select and train the optimal parameter values for reentrancy detection. We learn the model by optimizing binary cross-entropy loss. All experiments adopt the optimal gradient descent algorithm $Adam$\cite{kingma2014adam}. Our model searches for the learning rate $l_r$ in [0.0001, 0.0005, 0.001, 0.002, 0.005]. To prevent overfitting, we adjust the dropout rate $d_r$ searched in [0.2, 0.4, 0.6, 0.8]. The final parameters are set to: $l_r = 0.02$, $d_r = 0.2$, batch size $\beta = 64$, and vector dimension $vm = 300$.

\subsubsection{Evaluation Metrics and Experimental Results}
To evaluate the model's performance, we use metrics such as Accuracy (ACC), True Positive Rate (TPR), False Positive Rate (FPR), Precision (PRE), and F1-score. These metrics are calculated based on four scenarios: True Positives (TP), True Negatives (TN), False Positives (FP), and False Negatives (FN), as shown in Table.\ref{tab:sample_metrics}.

\begin{table}[htbp]
\centering
\caption{Table 4.5 Sample Metrics}
\label{tab:sample_metrics}
\begin{tabular}{|l|c|c|}
\hline
\multicolumn{1}{|c|}{\textbf{Actual Condition}} & \multicolumn{2}{c|}{\textbf{Predicted Condition}} \\ \hline
 & \textbf{Positive} & \textbf{Negative} \\ \hline
\textbf{Positive} & TP (True Positive) & FN (False Negative) \\ \hline
\textbf{Negative} & FP (False Positive) & TN (True Negative) \\ \hline
\end{tabular}
\end{table}

The formulas for the evaluation metrics are as follows:
\begin{equation}
    ACC = \frac{TP + TN}{TP + TN+FP+FN}
    \nonumber
\end{equation}

\begin{equation}
    TPR = \frac{TP}{TP+FN}
    \nonumber
\end{equation}

\begin{equation}
    FPR = \frac{FP}{FP+TN}
    \nonumber
\end{equation}

\begin{equation}
    PRE = \frac{TP}{TP+FP}
    \nonumber
\end{equation}

\begin{equation}
    F1 = \frac{2*PRE*TPR}{PRE+TPR}
    \nonumber
\end{equation}

For the BLSTM-ATT sequential model, we repeated the experiments 10 times to calculate the average performance, achieving favorable results. The BLSTM-ATT model achieved an F1-score of 88.26\% and an FPR of 8.57\%, indicating its ability to accurately identify reentrancy vulnerabilities. The high accuracy is likely due to the effectiveness of the BLSTM architecture and the attention mechanism. The BLSTM-ATT model not only captures long-term dependencies from both past and future contexts but also highlights key points through the attention mechanism.\par
\begin{figure*}[htbp]
    \centering
    \includegraphics[width=0.85\textwidth]{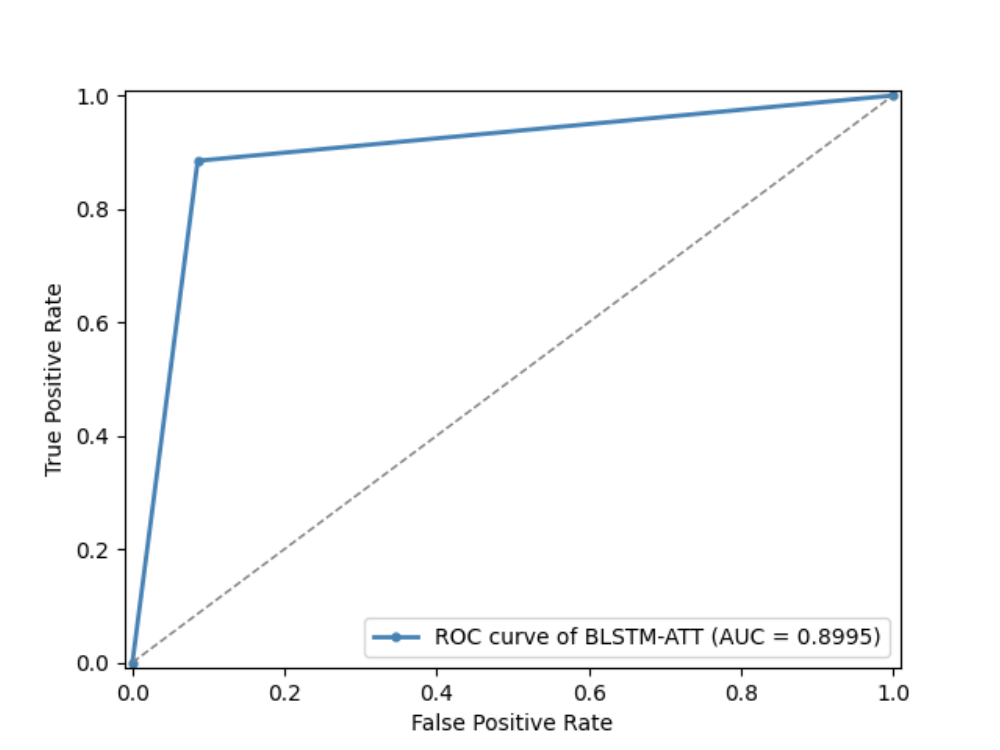}
    \caption{ROC Curve}
    \label{fig:7}
\end{figure*}

The performance of our sequential model was further analyzed using an ROC curve \cite{fawcett2006introduction}, as shown in Figure \ref{fig:7}. The ROC curve plots TPR on the y-axis and FPR on the x-axis and is commonly used to evaluate binary classifiers. The AUC for the BLSTM-ATT model is close to 90\%, indicating good detection performance.\par
From the results, we can conclude that deep learning-based detection methods, specifically sequential models, achieve the detection function effectively. This indicates that deep learning can be applied to vulnerability detection in smart contracts. Additionally, due to the semantic information capture of sequential models and the highlights of the attention mechanism, vulnerability detection in smart contracts can achieve high accuracy.\par
The model's good performance may be attributed to the contract snippets used, which discard useless information (such as code comments and blank lines) and capture key points (such as control flow dependencies, keywords, and semantic inheritance information). Through highly expressive contract snippets, our sequential model is well-adapted and trained to accurately identify vulnerabilities.
\vspace{-10pt}
\subsection{Privacy Protection}
With the growth of cloud computing, big data, and the Internet of Things, and the digital transformation of traditional industries, data volumes are growing exponentially, marking the arrival of the big data era. As a strategic resource, data plays an increasingly important role in national governance, economic growth, and national security, so research on data privacy and security is gaining more attention. Blockchain technology, known for its decentralization, traceability, and immutability, is widely used. However, these features come at the cost of disclosing certain information, such as transaction content being exposed due to data verifiability. To reduce privacy leakage risks, privacy protection on blockchains is essential, focusing mainly on identity privacy protection and transaction privacy protection \cite{li2021hybrid}.\par
We extract privacy requirements and threats from the network environment, transactions, and applications and conduct two types of analyses: The first analysis is based on the fundamental characteristics of blockchain, while the second provides a detailed description of various threats.
\subsubsection{Privacy Requirements of Blockchain}
\
\newline
In light of the characteristics of blockchain, we have conducted the following analyses:
\begin{itemize}
    \item In blockchain transactions, each block contains the hash value of the previous block, forming a chained structure. This means all transactions are traceable. Therefore, we need to make the connections between transactions invisible.
    \item Since all transaction information is stored in a public, global ledger, any participant can access and verify all data via public interfaces. Therefore, it is essential to protect identity information, i.e., the relationship between blockchain addresses and user identity information.
\end{itemize}
The threats to blockchain privacy protection primarily stem from the associations between user identity information and blockchain addresses, as well as the transaction records and the knowledge behind them stored in the blockchain.\par
\textbf{De-anonymization}: Malicious nodes can join the network without authorization and monitor communication data at the network layer. Attackers may link transaction information captured at the network layer with the originating node's IP address, thereby threatening user identity privacy.
\begin{itemize}
    \item Network Analysis: By monitoring data transmitted in a P2P network, attackers can obtain IP addresses when nodes broadcast transactions.
    \item Address Clustering: Users can divide the network into different address clusters. After labeling with data collection techniques, some addresses can be linked to the same user. While not easy to implement, these methods remain a potential threat and cannot be ignored.
    \item Denial of Service Attack: Malicious attackers exploit insufficiencies in network security measures, rendering normal service means unusable and causing machines or network resources to become unavailable.
    \item Sybil Attack: Malicious attackers use a small number of nodes to control multiple fake identities, thereby disrupting the balance of reputation systems in a P2P network.
\end{itemize}

\textbf{Transaction Pattern Analysis}: Other transaction flows to the public network can be analyzed statistically. For instance, transaction graph analysis can reveal overall transaction characteristics. AS-level deployment analysis involves recursively connecting to clients, requesting, and collecting the IP addresses of other peers to gather network information. This provides specific details about the scale, structure, and distribution of the core network.
\subsubsection{Identity Privacy Protection Methods}
\
\newline
Three common mechanisms for preserving anonymity in blockchain are: Coin Mixing\cite{ruffing2014coinshuffle}, Ring Signatures\cite{rivest2001leak}, and Non-interactive Zero-Knowledge Proofs\cite{gentry2009fully}.

\textbf{Coin Mixing}: Blockchain's transparency links transaction senders and receivers. Analyzing public data can reveal private info. A solution is to obscure transaction relationships using mixers, enhancing anonymity. \par
If an entity wants to send a message $M$ to another entity at address $R$, they encrypt $M$  with the recipient's public key $K_r$, attach address $R$, and then encrypt the result with the intermediary's public key $K_1$. The left side of the following expression shows the ciphertext transferred to the intermediary:
\begin{equation}
    K_1(r_0,K_R(r_1,M),R) \rightarrow K_R(r_1,M),R
\end{equation}
$\rightarrow$ represents transforming the initial ciphertext into a new ciphertext on the right. During this process, the intermediary decrypts the original ciphertext with their private key, then passes the sub-ciphertext to $R$, who decrypts it with their private key. Note that $r_1$ and $r_0$, as random numbers, ensure the message isn't transmitted multiple times.\par
The core idea of coin mixing is to resist transaction graph analysis without encrypting transaction content, thereby increasing the difficulty of attacks. By using intermediaries or spontaneous obfuscation to mix and transfer funds, attackers cannot directly obtain the sender-receiver correspondence in transactions, thus enhancing blockchain privacy protection.\par
Coin mixing techniques are categorized into centralized and decentralized approaches. Centralized methods use third-party nodes to obscure the link between transaction parties, making cryptocurrency flows harder to track. Examples include online wallets, dedicated mixing services, and multiple mixer overlays, which require no extra technical changes. Decentralized methods replace third-party nodes with a P2P protocol, combining multiple transactions into one with many inputs and outputs to hide associations.

\textbf{Ring Signature}: In a ring signature scheme, user $A$ selects a group of participants, including themselves, to form a ring {$user_0, user_1, ..., user_n$}. Each participant has a public key from a standard signature scheme (e.g., RSA, ECDSA). User $A$ signs a message using their private key $SK_A$ and all the public keys {$PK_0, PK_1, ..., PK_n$} of the ring members. The verifier can confirm that the message was signed by one member of the group but cannot identify the actual signer. This provides complete anonymity for the signer. Ring signatures hide the signer's identity among the public keys of the ring, with no centralized authority or administrator involved.

\textbf{Non-interactive Zero-Knowledge Proof(NIZK)}: A zero-knowledge proof (ZKP) \cite{wang2022non} is a cryptographic method that allows one to prove a statement without revealing any additional information. NIZK differs from ZKP in that it requires no interaction between the prover and verifier, making it suitable for anonymous and distributed message verification in blockchain systems.
A formal definition of a Non-Interactive Zero-Knowledge (NIZK) \cite{grilo2025distributed} proof system is as follows: Let ${(P, V)}$ denote a pair of probabilistic polynomial-time algorithms acting as the prover and verifier, respectively. For a language ${\mathcal{L} \subseteq \mathsf{NP}}$ (with a security parameter $k$), the tuple ${(P, V)}$  is called an NIZK proof system for $\mathcal{L}$ if it satisfies the following properties:
\begin{itemize}
    \item Integrity: For any input $x \in \mathcal{L}$, its witness $w$ , and polynomial $p()$, the following must satisfied:
    \begin{equation}
        P_r[V(R,x,P(R,x,w))= 1] \ge 1 - \frac{1}{p(|x|)}
    \end{equation}
    \item Soundness: For any input $x \notin \mathcal{L}$,any probabilistic polynomial-time algorithm $p^*$, and any polynomial $P()$, the following must be satisfied: 
    \begin{equation}
        P_r[V(R,x,P^*(R,x))=1] < \frac{1}{P(|x|)}
    \end{equation}
    \item Zero-Knowledge: For any $x \in \mathcal{L}$,  its witness $w$, there exists a probabilistic polynomial-time simulator $S$ such that the following distributions are computationally indistinguishable:
    \begin{equation}
        \{R,x,P(R,x,w)\} \approx \{R,x,\pi\} \leftarrow S(x)
    \end{equation}
    This means that all information obtained by the verifier during interaction with the prover can also be computed by a probabilistic polynomial-time simulator. Note that $R$ is a public random reference string.
\end{itemize} 
\subsubsection{Transaction Privacy Protection Methods}
\
\newline
Homomorphic encryption systems (HC) allow computations on ciphertexts without decrypting them. This means that operations on encrypted data yield results that match those of the same operations on plaintext. This enables tasks like querying encrypted data without exposing it, thus enhancing privacy when data is outsourced or stored with third parties.

Consider a scenario where Party $A$ holds values $(x_1,x_2,...x_n)$,and Party $B$ holds a function $f()$. Both want to compute $f(x_1,x_2,...x_n)$ without disclosing the values or the function's details. In a homomorphic encryption system, $A$ encrypts the inputs $\{E(x_1),...,E(x_n)\}$ and sends them to $B$. $B$ performs the computation on the ciphertexts, randomizes the result, and sends it back to $A$. Upon decryption, $A$ securely obtains $f(x_1,x_2,...x_n)$.

Homomorphic encryption systems, as black-box operations, take $n$ ciphertexts and operations as input and output the encrypted result of the corresponding operation on the original data. This feature makes them ideal for securely updating transaction amounts and other data in blockchains. Typical homomorphic encryption schemes for blockchain privacy protection include the Pedersen Commitment Scheme\cite{pedersen1991non} and the Paillier System \cite{paillier1999public}.
\section{Future Research Directions}
From the above chapters, we understand common privacy protection methods. This section outlines future research directions in this field.
\begin{itemize}
    \item Scalability: Coin mixing causes extra waiting delays, and complex cryptographic primitives often lead to significant computational and communication overhead. These high costs limit the scalability of anonymity. Thus, one possible direction is to solve the combinatorial optimization problem between existing or new cryptographic primitives and their potential configurations.
    \item Enhancing Privacy under Weaker Assumptions:
Strengthen privacy protection in scenarios with minimal or no trust assumptions.
    \item Compatibility: A major challenge is ensuring compatibility between privacy protection methods and account architectures, such as Ethereum's account system, which maintains addresses as a global state. Ethereum, being the most widely used platform due to its built-in Turing-complete programming language, is considered ideal for DAPP development. However, integrating privacy protection with its account architecture remains a significant challenge.
    \item Privacy Protection and Regulatable Control:
Blockchain's decentralized and trustless nature has driven its widespread adoption, with privacy protection safeguarding user data. However, these technologies can be exploited for illegal activities like money laundering. Thus, blockchain activities need regulation by a trusted institution to prevent misuse while still protecting users' sensitive data.
\end{itemize}

\section{Conclusion}
This paper focuses on the privacy and security issues of blockchain-based decentralized applications (DApp). Through data behavior analysis of DApp, we reveal their development status and conduct security analyses, supporting subsequent chapters. We also detect reentrancy vulnerability attacks in smart contracts using the BLSTM-ATT model, enabling source code-level attack detection. Furthermore, we examine privacy threats in blockchain and discuss cryptographic defenses like identity and transaction privacy protection. Finally, we summarize existing privacy protection methods and outline future research challenges in this field.
Research on the privacy and security of blockchain-based decentralized applications is a vast topic. This paper has only scratched the surface by analyzing and discussing some basic knowledge, security threats, and corresponding methods. For the underlying blockchain technology of DApp, its security issues are complex and directly impact DApp security. Challenges such as data security at the data layer and consensus algorithm security at the consensus layer warrant further exploration. Additionally, smart contracts, a unique component of DApp, face not only reentrancy attacks but also other threats like short-address attacks and code injection, which will remain research hotspots.

\vspace{10pt}

\bibliographystyle{apalike} 


\end{document}